\documentclass[sigconf]{acmart}

\usepackage{amsmath,amsthm}
\usepackage{booktabs}
\usepackage[ruled]{algorithm2e}
\usepackage{tabularx}
\usepackage{multirow}
\usepackage{enumitem}

\setlist{leftmargin=3.5mm}
\theoremstyle{definition}
\newtheorem{definition}{Definition}

\AtBeginDocument{%
  \providecommand\BibTeX{{%
    \normalfont B\kern-0.5em{\scshape i\kern-0.25em b}\kern-0.8em\TeX}}}

\makeatletter
\renewcommand{\@copyrightpermission}{}
\makeatother
\copyrightyear{2020}
\acmYear{2020}
\acmDOI{10.1145/1122445.1122456}

\acmConference[Cleveland]{2020: The 26th ACM SIGKDD International Conference on Knowledge Discovery \& Data Mining}{OH}{2020}
\acmBooktitle{KDD '20: The 26th ACM SIGKDD International Conference on Knowledge Discovery \& Data Mining, August 22--27, 2020, San Diego, CA}
\acmPrice{15.00}
\acmISBN{978-1-4503-XXXX-X/18/06}

\begin{document}

\title{\fontsize{14}{20}\selectfont \textbf{$\alpha$-Satellite}: An AI-driven System and Benchmark Datasets for \\Hierarchical Community-level Risk Assessment to Help Combat COVID-19}

\author{Yanfang Ye$^{*1}$, Shifu Hou$^{1}$, Yujie Fan$^{1}$, Yiyue Qian$^{1}$}
\author{Yiming Zhang$^{1}$, Shiyu Sun$^{1}$, Qian Peng$^{1}$, Kenneth Laparo$^{2}$}
\affiliation{
	\institution{$^1$Department of Computer and Data Sciences, Case Western Reserve University, OH, USA}
	\institution{$^2$Department of Electrical, Computer, and Systems Engineering, Case Western Reserve University, OH, USA}
}
\email{{yanfang.ye, sxh1055, yxf370, yxq250, xyz2092, sxs2293, qxp36, kal4}@case.edu}


\begin{abstract}
	
The novel coronavirus and its deadly outbreak have posed grand challenges to human society: as of March 26, 2020, there have been 85,377 confirmed cases and 1,293 reported deaths in the United States; and the World Health Organization (WHO) characterized coronavirus disease (COVID-19) - which has infected more than 531,000 people with more than 24,000 deaths in at least 171 countries - a global pandemic. A growing number of areas reporting local sub-national community transmission would represent a significant turn for the worse in the battle against the novel coronavirus, which points to \textit{an urgent need} for expanded surveillance so we can better understand the spread of COVID-19 and thus better respond with actionable strategies for community mitigation. By advancing capabilities of artificial intelligence (AI) and leveraging the large-scale and real-time data generated from heterogeneous sources (e.g., disease related data from official public health organizations, demographic data, mobility data, and user geneated data from social media), in this work, we propose and develop an AI-driven system (named $\alpha$\textit{-Satellite}), \textit{as an initial offering}, to provide hierarchical community-level risk assessment to assist with the development of strategies for combating the fast evolving COVID-19 pandemic. More specifically, given a specific location (either user input or automatic positioning), the developed system will automatically provide risk indexes associated with it in a hierarchical manner (e.g., state, county, city, specific location) to enable individuals to select appropriate actions for protection while minimizing disruptions to daily life to the extent possible. The developed system and the generated benchmark datasets have been made publicly accessible through our website\footnote{https://COVID-19.yes-lab.org/}.	The system description and disclaimer are also available in our website.

\end{abstract}


\maketitle
\fancyhead{}

\newpage
\section{Introduction} \label{intro}

\begin{figure}[htbp!]
	\centering
	\includegraphics[width=0.92\linewidth]{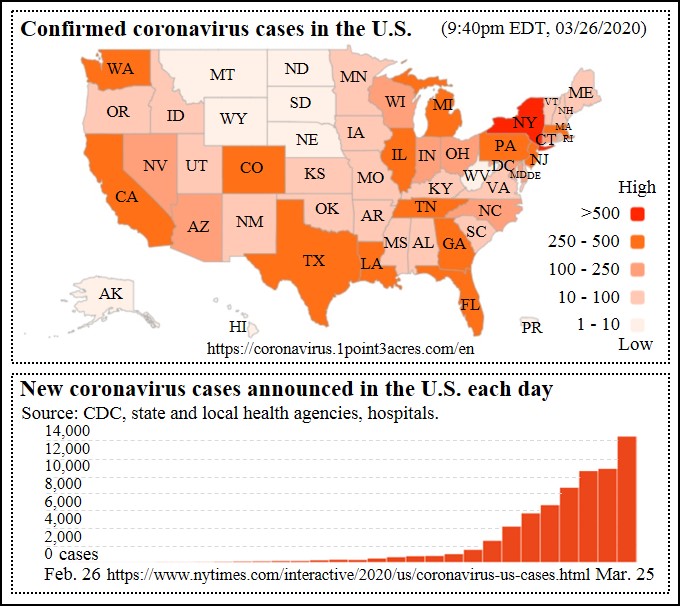}
	\vspace{-0.25cm}
	\caption{COVID-19 in the U.S. (by 2:59pm EDT, 03/26/20).} \label{fig:intro1}
\end{figure}

Coronavirus disease (COVID-19) \cite{COVID-19} is an infectious disease caused by a new virus that had not been previously identified in humans; this respiratory illness (with symptoms such as a cough, fever and pneumonia) was first identified during an investigation into an outbreak in Wuhan, China in December 2019 and is now rapidly spreading in the U.S. and globally. The novel coronavirus and its deadly outbreak have posed grand challenges to human society. As of March 26, 2020, there have been 85,377 confirmed cases and 1,293 reported deaths in the U.S. (Figure~\ref{fig:intro1}); and the WHO characterized COVID-19 - which has infected more than 531,000 people with more than 24,000 deaths in at least 171 countries - a global pandemic. 

\vspace{-0.2cm}
\vspace{-0.3cm}

\begin{figure}[htbp!]
	\centering
	\includegraphics[width=1\linewidth]{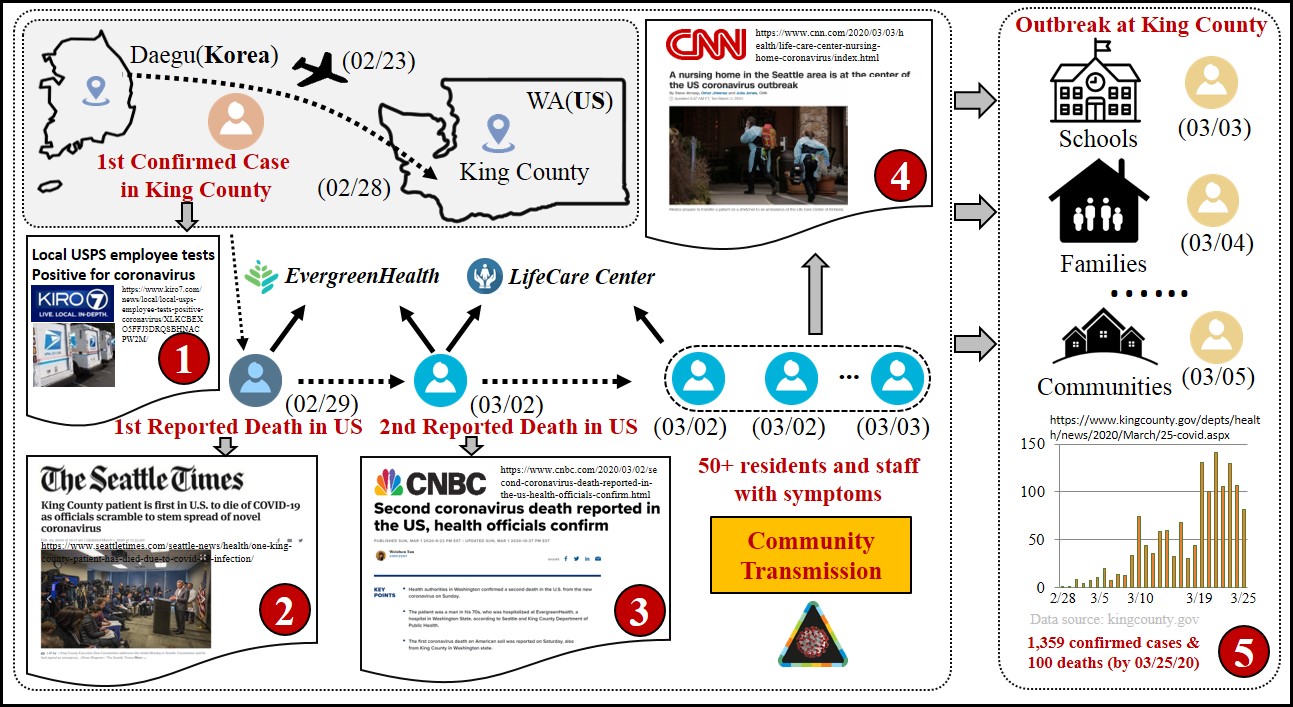}
	\vspace{-0.6cm}
	\caption{An example of community transmission.} \label{fig:king}
\end{figure}

It is believed that the novel virus which causes COVID-19 emerged from an animal source, but it is now rapidly spreading from person-to-person through various forms of contact. According to the Centers for Disease Control and Prevention (CDC) \cite{CDC-COVID19}, the coronavirus seems to be spreading easily and sustainably in the community - i.e., \textit{community transmission} which means people have been infected with the virus in an area, including some who are not sure how or where they became infected. An example of community transmission that caused the outbreak of COVID-19 in King County at Washington State (WA) is shown in Figure~\ref{fig:king}. The challenge with community transmission is that carriers are often asymptomatic and unaware that they are infected and through their movements within the community they spread the disease. According to the CDC, before a vaccine or drug becomes widely available (i.e., this is the case for COVID-19 by far), \textit{community mitigation}, which is a set of actions that persons and communities can take to help slow the spread of respiratory virus infections, is the most readily available interventions to help slow transmission of the virus in communities \cite{CDC-Mitigation}. A growing number of areas reporting local sub-national community transmission would represent a significant turn for the worse in the battle against the novel coronavirus, which points to \textbf{an urgent need} for expanded surveillance so we can better understand the spread of COVID-19 and thus better respond with actionable strategies for community mitigation. 

\vspace{-0.2cm}
\vspace{-0.3cm}

Unlike the 1918 influenza pandemic \cite{CDC-1918Pandemic} where the global scope and devastating impacts were only determined well after the fact, COVID-19 history is being written daily, if not hourly, and if the right types of data can be acquired and analyzed there is the potential to improve self awareness of the risk to the population and develop proactive (rather than reactive) interventions that can halt the exponential growth in the disease that is currently being observed. Realizing the true potential of real-time surveillance, with this opportunity comes the challenge: the available data are uncertain and incomplete while we need to provide mitigation strategies objectively with caution and rigor (i.e., enable people to select appropriate actions to protect themselves at increased risk of COVID-19 while minimize disruptions to daily life to the extent possible). 

To address the above challenge, leveraging our long-term and successful experiences in combating and mitigating widespread malware attacks using AI-driven techniques \cite{Ye_AiDroid, Ye_CIKM2019, Ye_AICS, Ye_KDD2018, Ye_CSUR:2017, Ye_KIS:2017, hou2017hindroid, Chen_ACSAC:2017, Chen_EISIC:2017, Fan_ESWA:2016, Ye_KDD:2011, Ye_KDD:2010, Ye_SMC:2010, Ye_KDD:2009, Ye_JCV:2008, Ye_KDD:2007}, in this work, we propose to design and develop \textit{an AI-driven system to provide hierarchical community-level risk assessment at the first attempt} to help combat the fast evolving COVID-19 pandemic, by using the large-scale and real-time data generated from heterogeneous sources. In our developed system (named $\alpha$\textit{-Satellite}), we first develop a set of tools to collect and preprocess the large-scale and real-time pandemic related data from multiple sources, including disease related data from official public health organizations, demographic data, mobility data, and user generated data from social media; and then we devise advanced AI-driven techniques to provide hierarchical community-level risk assessment to enable actionable strategies for community mitigation. More specifically, given a specific location (either user input or automatic positioning), the developed system will automatically provide risk indexes associated with it in a hierarchical manner (e.g., state, county, city, specific location) to enable people to select appropriate actions for protection while minimizing disruptions to daily life.

\begin{figure*}[th]
	\centering
	\includegraphics[width=1.0\linewidth]{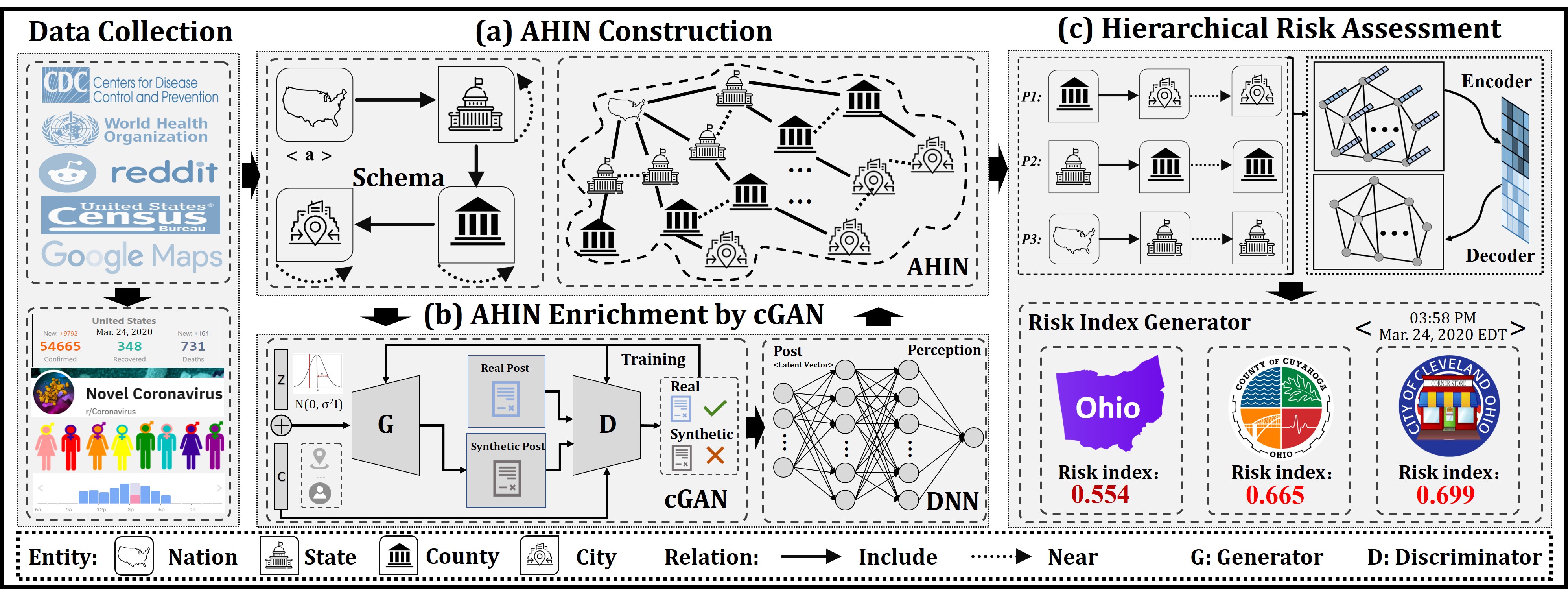}
	\vspace{-0.5cm}
	\caption{System architecture of $\alpha$\textit{-Satellite} (i.e., an AI-driven system for hierarchical community-level risk assessment). In $\alpha$\textit{-Satellite}, \textit{(a)} we first construct an AHIN to model the collected large-scale and real-time pandemic related data in a comprehensive way; \textit{(b)} based on the constructed AHIN, we then exploit the cGANs to gain the public perception towards COVID-19 in an given area; \textit{(c)} we finally utilize meta-path based schemes to model both vertical and horizontal information associated with a given area, and devise heterogeneous GAE to aggregate information from its neighborhood areas to estimate the risk of the given area in a hierarchical manner.}
	\label{fig:sys}
\end{figure*}

The framework of our proposed and developed system is shown in Figure \ref{fig:sys}. In the system of $\alpha$\textit{-Satellite}, (1) we first construct an attributed heterogeneous information network (AHIN) to model the collected large-scale and real-time pandemic related data in a comprehensive way; (2) based on the constructed AHIN, to address the challenge of limited data that might be available for learning (e.g., social media data to learn public perceptions towards COVID-19 in a given area might not be sufficient), we then exploit the conditional generative adversarial nets (cGANs) to gain the public perceptions towards COVID-19 in each given area; and finally (3) we utilize meta-path based schemes to model both vertical and horizontal information associated with a given area, and devise a novel heterogeneous graph auto-encoder (GAE) to aggregate information from its neighborhood areas to estimate the risk of the given area in a hierarchical manner. The developed system $\alpha$\textit{-Satellite} and the generated benchmark datasets have been made publicly accessible through our website.

\section{Related Work}\label{sec:RelatedWork}

There have been several works on using AI and machine learning techniques to help combat COVID-19: in the biomedical domain, \cite{xu2020deep,chen2020deep,wang2020deep,song2020deep,randhawa2020machine} use deep learning methods for COVID-19 pneumonia diagnosis and genome study; while \cite{yan2020prediction, shi2020deep} develop learning-based models to predict severity and survival for patients. Another research direction is to utilize public accessible data to help the estimation of infection cases or forecast the COVID-19 outbreak  \cite{hu2020artificial,hermanowicz2020forecasting,jahanbin2020using,majumder2020early,song2020epidemiological,rao2020identification,zhu2020host}. However, most of these existing works mainly focus on Wuhan China; the studies of using computational models to combat COVID-19 in the U.S. are scarce and there has no work on community-level risk assessment to assist with community mitigation by far. \textbf{\textit{To meet this urgent need and to bridge the research gap}}, in this work, by advancing capabilities of AI and leveraging the large-scale and real-time data generated from heterogeneous sources, we propose and develop an AI-driven system, named $\alpha$\textit{-Satellite}, to provide hierarchical community-level risk assessment \textit{at the first attempt} to help combat the deadly and fast evolving COVID-19 pandemic.

\section{Proposed Method}\label{sec:problemdefinition}

In this section, we will introduce our proposed method integrated in the system of $\alpha$\textit{-Satellite} to automatically provide hierarchical community-level risk assessment related to COVID-19 in detail.

\vspace{-0.2cm}
\subsection{Data obtained from Heterogeneous Sources} \label{data}

Realizing the true potential of real-time surveillance requires identifying the proper data sources, based on which we can devise models to extract meaningful and actionable information for community mitigation. Since relying on a single data source for estimation and prediction often results in unsatisfactory performance, we develop a set of crawling tools and preprocessing methods to collect and parse the large-scale and real-time pandemic related data from multiple sources, which include the followings. 

\vspace{-0.1cm}
\begin{itemize}
	\item \textit{Disease related data}. We collect the up-to-date county-based coronavirus related data including the numbers of confirmed cases, new cases, deaths and the fatality rate, from i) official public health organizations such as WHO, CDC, and county government websites, and ii) digital media with real-time updates of COVID-19 (e.g., 1point3acres\footnote{https://coronavirus.1point3acres.com/en}). The collected up-to-date county-based COVID-19 related statistical data can be an important element for risk assessment of an associated area.

	\item \textit{Demographic data}. The United States Census Bureau\footnote{https://www.census.gov/quickfacts/fact/table/US/PST045219} provides the demographic data including basic population, business, and geography statistics for all states and counties, and for cities and towns with more than 5,000 people. The demographic information will contribute to the risk assessment of an associated area: for example, as older adults may be at higher risk for more serious complications from COVID-19 \cite{CDC-olderpeople,surveillances2020epidemiological}, the age distribution of a given area can be considered as an important input. In this work, given a specific area, we mainly consider the associated demographic data including the estimated population, population density (e.g., number of people per square mile), age and gender distributions. 
	
	\item \textit{Mobility data}. Given a specific area (either user input or automatic positioning), a mobility measure that estimates how busy the area is in terms of traffic density will be retained from location service providers (i.e., Google maps).  
	
	\item \textit{User generated data from social media}. As users in social media are likely to discuss and share their experiences of COVID-19, the data from social media may contribute complementary knowledge such as public perceptions towards COVID-19 in the area they associate with. In this work, we initialize our efforts with the focus on Reddit, as it provides the platform for scientific discussion of dynamic policies, announcements, symptoms and events of COVID-19. In particular, we consider i) three subreddits with general discussion (i.e., r/Coronavirus\footnote{https://www.reddit.com/r/Coronavirus/}, r/COVID19\footnote{https://www.reddit.com/r/COVID19/} and r/CoronavirusUS\footnote{https://www.reddit.com/r/CoronavirusUS/}); ii) four region-based subreddits, which are r/CoronavirusMidwest, r/CoronavirusSouth, r/CoronavirusSouthEast and r/CoronavirusWest; and iii) 48 state-based subreddits (i.e., Washington, D.C. and 47 states). To analyze public perceptions towards COVID-19 for a given area (\textit{note that} all users are anonymized for analysis using hash values of usernames), we first exploit Stanford Named Entity Recognizer \cite{finkel2005incorporating} to extract the location-based information (e.g., county, city), and then utilize tools such as NLTK \cite{bird2009natural} to conduct sentiment analysis (i.e., positive, neutral or negative). More specifically, positive denotes well aware of COVID-19, while negative indicates less aware of COVID-19. For example, with the analysis of the post by a user (with hash value of ``CF***6'') in subreddit of r/CoronaVirusPA on March 14, 2020: \textit{``I live in Montgomery County, PA and everyone here is acting like there's nothing going on.''}, the location-related information of Montgomery county and Pennsylvania state (i.e., PA) can be extracted, and a user's perception towards COVID-19 in Montgomery county at PA can be learned (i.e., negative indicating less aware of COVID-19). Such automatically extracted knowledge will be incorporated into the risk assessment of the related area; meanwhile, it can also provide important information to help inform and educate about the science of coronavirus transmission and prevention.
	
\end{itemize}
\vspace{-0.2cm}

\subsection{AHIN built from Collected Data} \label{FeatureExtraction}

To comprehensively describe a given area for its risk assessment related to COVID-19, based on the data collected from multiple sources above, we consider and extract higher-level semantics as well as social and behavioral information within the communities.

\noindent \textbf{Attributed Features.} Based on the collected data above, we further extract the corresponding attributed features.

\vspace{-0.1cm}
\begin{itemize}
	
	\item \textit{A1: disease related feature.} For a given area, its related COVID-19 pandemic data will be extracted including the numbers of confirmed cases, new cases, deaths and the fatality rate, which is represented by a numeric feature vector $\mathbf{a}_{1}$. For example, as of March 22, 2020, the Cuyahoga County at Ohio State (OH) has had 125 confirmed cases, 33 new cases, 1 death and 0.8\% fatality rate, which can be represented as $\mathbf{a_1}=<125, 33, 1, 0.008>$.

	\item \textit{A2: demographic feature.} Given a specific area, we obtain its associated city's (or town's) demographic data from the United States Census Bureau, including the estimated population, population density (i.e., number of people per square mile), age distribution (i.e., percentage of people over 65 year-old) and gender distribution (i.e., percentage of females). For example, to assist with the risk assessment of the area of Euclid Ave in Cleveland at OH, the obtained demographic data associated with it are: Cleveland with population of 383793, population density of 5107, 13.5\% people over 65 year-old, and 51.8\% females, which will be represented as $\mathbf{a}_2=<383793, 5107, 0.135, 0.518>$.
	
	\item \textit{A3: mobility feature.} Given a specific area, a mobility measure that estimates how busy the area is in terms of traffic density will be obtained from Google maps, which will represented by five degree levels (i.e., [1,5], the larger number the busier). 
	
	\item \textit{A4: representation of public perception.} After performing the automatic sentiment analysis based on the collected posts associated with a given area from Reddit, the public perceptions towards COVID-19 in this area will be represented by a normalized value (i.e., [0,1]) indicated the awareness of COVID-19 (i.e., the larger value the more aware). For the previous example of the Reddit post of \textit{``I live in Montgomery County, PA and everyone here is acting like there's nothing going on.''}, a related perception towards COVID-19 in Montgomery County at PA will be formulated as a numeric vale of $0.220$, denoting people in this area were less aware of COVID-19 on March 14, 2020.

\end{itemize}    
\vspace{-0.1cm}

After extracting the above features, we concatenate them as a normalized attributed feature vector \textit{A} attached to each given area for representation, i.e., $A=A_1\oplus A_2\oplus A_3\oplus A_4$. Note that we zero-pad the ones in the elements when the data are not available.

\noindent \textbf{Relation-based Features.} Besides the above extracted attributed features, we also consider the rich relations among different areas.

\vspace{-0.1cm}
\begin{itemize}	
	\item \textit{R1: administrative affiliation.} According to the severity of COVID-19, available resources and impacts to the residents, different states may have different policies, actionable strategies and orders with responses to COVID-19. Therefore, given an area, we accordingly extract its administrative affiliation in a hierarchical manner. Particularly, we acquire the \textit{state-include-county} and \textit{county-include-city} relations from City-to-County Finder\footnote{http://www.stats.indiana.edu/uspr/a/place\_frame.html}.
	
	\item \textit{R2: geospatial relation.} We also consider the geospatial relations between a given area and its neighborhood areas. More specifically, given an area, we retain its $k$-nearest neighbors at the same hierarchical level by calculating the euclidean distances based on their global positioning system (GPS) coordinates obtained from Google maps and Wikipedia\footnote{https://en.wikipedia.org/wiki/User:Michael\_J/County\_table}.	 
	
\end{itemize}
\vspace{-0.1cm}

\noindent \textbf{AHIN Construction.} Given the rich semantics and complex relations extracted above, it is important to model them in a proper way so that different relations can be better and easier handled. To solve this problem, we introduce AHIN to model them, which is able to be composed of different types of entities associated with attributed features and different types of relations. We first present the concepts related to AHIN below.

\vspace{-0.1cm}
\begin{definition}
	\textbf{\textit{Attributed Heterogeneous Information Network (AHIN)}} \cite{li2017semi}: Let $\mathcal{T}=\{T_1,...,T_m\}$ be a set of $m$ entity types, $\mathcal{X}_i$ be the set of entities of type $T_i$ and $A_i$ be the set of attributes defined for entities of type $T_i$. An AHIN is defined as a graph ${\mathcal G} = (\mathcal V, \mathcal E, \mathcal A)$ with an entity type mapping $\phi$: ${\mathcal V} \to \mathcal T$ and a relation type mapping $\psi$: ${\mathcal E} \to \mathcal R$, where $\mathcal {V}=\bigcup_{i=1}^m\mathcal{X}_i$ denotes the entity set and ${\mathcal E}$ is the relation set, $\mathcal T$ denotes the entity type set and $\mathcal R$ is the relation type set, $\mathcal{A}=\bigcup_{i=1}^mA_i$, and $|\mathcal T|+|\mathcal R|>2$. \textbf{\textit{Network Schema}} \cite{li2017semi}: The network schema of an AHIN $\mathcal G$ is a meta-template for $\mathcal G$, denoted as a directed graph $\mathcal T_{\mathcal G} = (\mathcal T, \mathcal R)$ with nodes as entity types from $\mathcal T$ and edges as relation types from $\mathcal R$.
\end{definition}
\vspace{-0.1cm}

\begin{figure}[htbp!]
	\centering
	\includegraphics[width=0.96\linewidth]{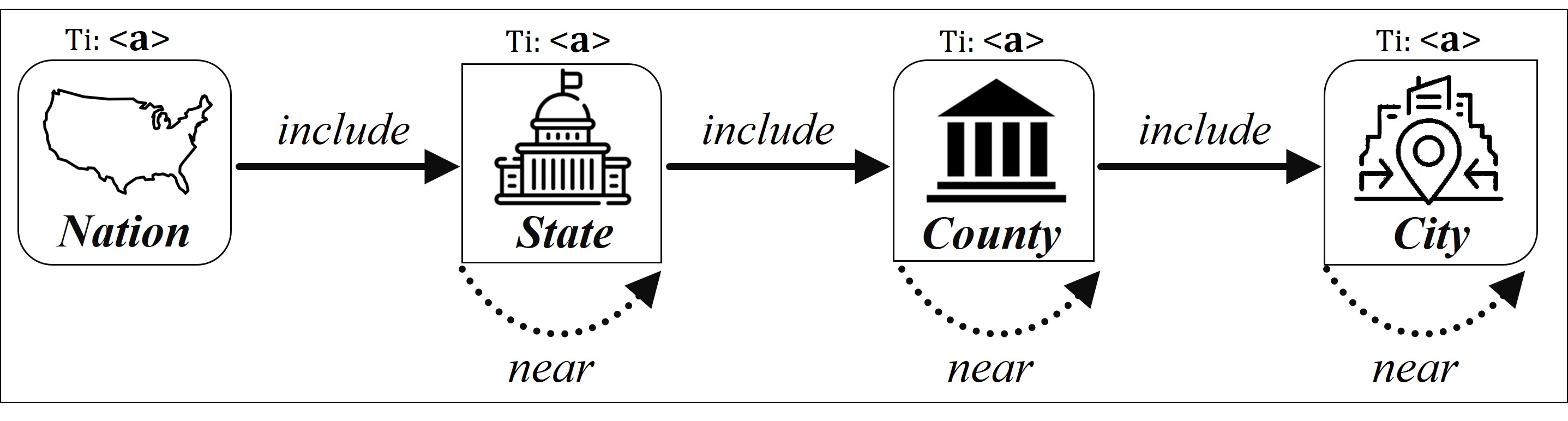}
	\vspace{-0.3cm}
	\caption{Network schema of AHIN.}
	\label{fig:schema}
\end{figure}

In this work, we have four types of entities (i.e., nation, state, county and city, $|\mathcal T|=4$), two types of relations (i.e., \textit{R1} and \textit{R2}, $|\mathcal R|=2$), and each entity is attached with an attributed feature vector as described above. Based on the definitions, the network schema of AHIN in our case is shown in Figure \ref{fig:schema}.

\vspace{-0.2cm}
\vspace{-0.3cm}

\vspace{-0.2cm}
\subsection{AHIN Enrichment by cGAN}

Although the constructed AHIN can model the complex and rich relations among different entities attached with attributed features, there faces a challenge that there might be missing values of attributed features attached to the entities in the AHIN because of limited data that might be available for learning. More specifically, given an area, there may not be sufficient social media data (i.e., Reddit data in this work) to learn the public perceptions towards COVID-19 in this area. For example, for the state of Montana, as of March 22, 2020, in its corresponding subreddit r/CoronavirusMontana, there only have been 12 posts by seven users discussing the virus. To address this issue, we propose to exploit cGANs \cite{mirza2014conditional} for synthetic (virtual) social media user data generation for public perception learning to enrich the AHIN.

\begin{figure}[htbp!]
	\centering
	\includegraphics[width=1\linewidth]{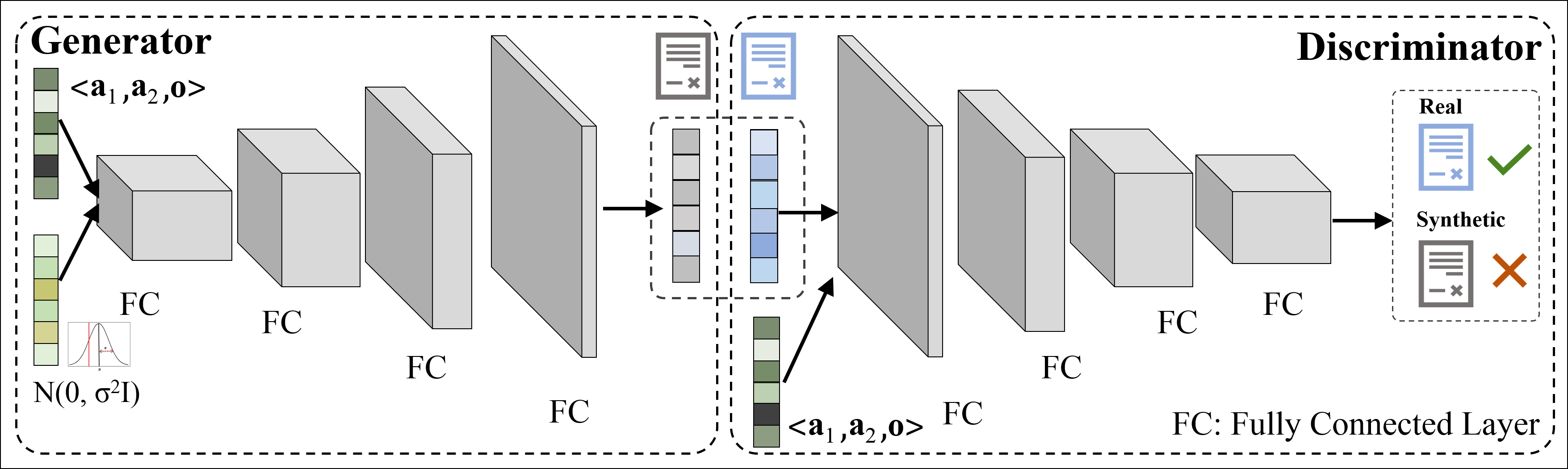}
	\vspace{-0.5cm}
	\caption{cGAN for synthetic post latent vector generation.}
	\label{fig:cGAN}
\end{figure}

Different from traditional GANs \cite{goodfellow2014generative}, a cGAN is a conditional model extended from GANs, where both the generator and discriminator are conditioned on some extra information. In our case, we propose to exploit cGAN to generate the synthetic posts for those areas where the data are not available. In our designed cGAN, given an area where Reddit data are not available, the condition composes of three parts: the disease related feature vector in this area $\mathbf{a}_1$, its related demographic feature vector $\mathbf{a}_2$ and its GPS coordinate denoted as $\mathbf{o}$. As shown in Figure \ref{fig:cGAN}, the generator in the devised cGAN aims to incorporate the prior noise $p_z(\mathbf z)$, with the conditions of $\mathbf{a}_1$, $\mathbf{a}_2$ and $\mathbf o$ as the inputs to generate the synthetic posts represented by latent vectors; while in the discriminator, real post representations obtained by using \textit{doc2vec} \cite{le2014distributed} or generated synthetic post latent vectors along with $\mathbf{a}_1$, $\mathbf{a}_2$ and $\mathbf o$ are fed to a discriminative function. Both generator and discriminator could be a non-linear mapping function, such as a multi-layer perceptron (MLP). The generator and discriminator play the adversarial minimax game formulated as the following minimax problem: 
\begin{equation}\label{key}
\begin{aligned}
\min_G\max_D V(D,G)&=\mathbb{E}_{\mathbf p\sim p_{data}(\mathbf p)}[\log D(\mathbf p|\mathbf{a}_1, \mathbf{a}_2,\mathbf o)]\\
&+\mathbb{E}_{\mathbf z\sim p_{z}(\mathbf z)}[\log (1-D(G(\mathbf z|\mathbf{a}_1, \mathbf{a}_2,\mathbf o)))].
\end{aligned}
\end{equation}
The generator and discriminator are trained simultaneously: adjusting parameters for generator to minimize $\log (1-D(G(\mathbf z|\mathbf{a}_1, \mathbf{a}_2,\mathbf o)))$ while adjusting parameters for discriminator to maximize the probability of assigning the correct labels to both training examples and generated samples.

\vspace{-0.2cm}
\vspace{-0.3cm}

After applying cGAN for synthetic post latent vector generation, we further exploit deep neural network (DNN) to learn the public perceptions towards COVID-19 in this area. More specifically, we first use \textit{doc2vec} to obtain the representations of real posts collected from Reddit and feed them to train the DNN model; and then given a generated synthetic post latent vector, we use the trained model to gain its related perception (i.e., awareness of COVID-19).

\vspace{-0.1cm}
\subsection{Hierarchical Risk Assessment}\label{sec:HGNN}

\noindent \textbf{Meta-path Expression.} To assist with the risk assessment of a given area related to the fast evolving COVID-19, it might not be sufficient if only considering its vertical information (e.g., its related city, county or state's responses, strategies and policies); the horizontal information (i.e., information from its neighborhood areas) will also be important inputs. To comprehensively integrate both vertical and horizontal information, we propose to exploit the concept of meta-path \cite{sun2011pathsim} to formulate the relatedness among different areas in the constructed AHIN.

\vspace{-0.1cm}
\begin{definition}
	\textbf{\textit{Meta-path.}} A meta-path $ \mathcal{P} $ is a path defined on the network schema $\mathcal T_{\mathcal G}=(\mathcal{T, R}) $, and is denoted in the form of $ T_{1} \xrightarrow{R_{1}} T_{2} \xrightarrow{R_{2}} ... \xrightarrow{R_{L}} T_{L+1} $, which defines a composite relation $R = R_1 \cdot R_2 \cdot \ldots \cdot R_L$ between types $T_1$ and $T_{L+1}$, where $\cdot$ denotes relation composition operator, and $L$ is the length of $\mathcal P$. 
\end{definition}
\vspace{-0.2cm}

\vspace{-0.2cm}
\begin{figure}[htbp!]
	\centering
	\includegraphics[width=1\linewidth]{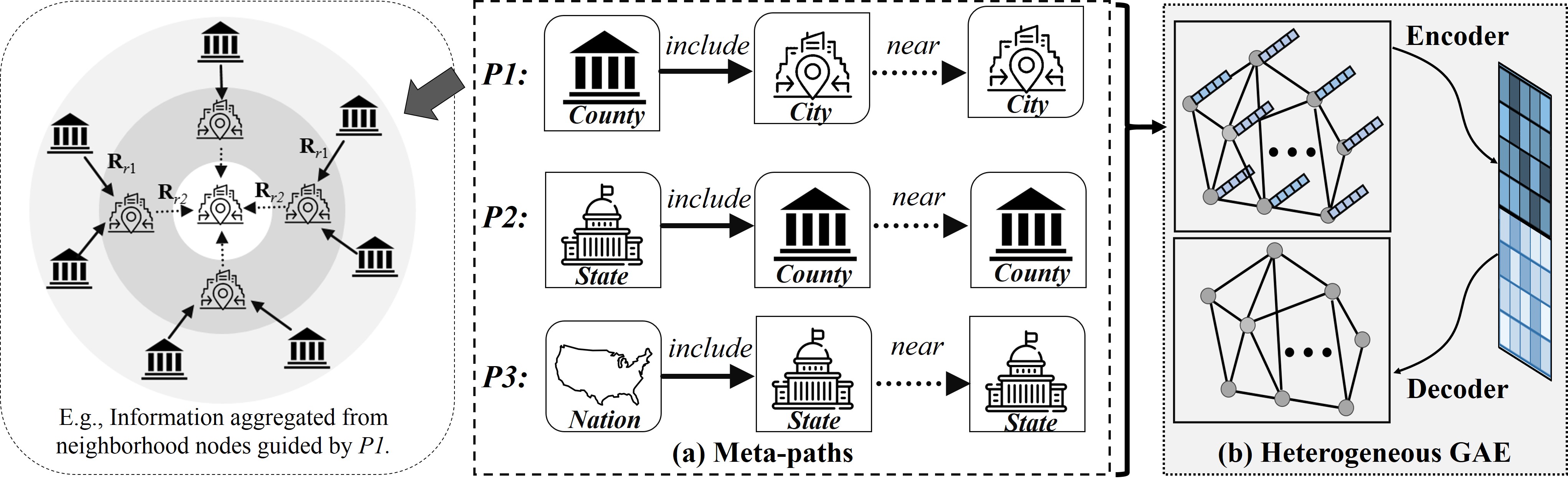}
	\vspace{-0.7cm}
	\caption{Meta-paths and heterogeneous GAE.}
	\label{fig:HGNN}
\end{figure}
\vspace{-0.2cm}

Figure \ref{fig:HGNN}.(a) shows our designed meta-paths (i.e., \textit{P1-P3}). For example, \textit{P1} of $county \xrightarrow{include} city \xrightarrow{near} city$ denotes that, to assess the risk of a specific city, we not only consider the city itself, but also the information from its related county and nearby cities.

\noindent \textbf{Heterogeneous Graph Auto-encoder.} Given a node (i.e., area) in the constructed AHIN, guided by its corresponding meta-path scheme (i.e., city level guided by \textit{P1}, county level guided by \textit{P2}, and state level guided by \textit{P3}), to aggregate the information propagated from its neighborhood nodes, we propose a heterogeneous graph auto-encoder (GAE) model to achieve this goal. The designed heterogeneous GAE model consists of an encoder and a decoder: the encoder aims at encoding meta-path based propagation to a latent representation, and the decoder will reconstruct the topological information from the representation. 

\noindent \textit{Encoder.} We here exploit attentive mechanism \cite{velivckovic2017graph,fan2019metapath,wang2019kgat} to devise the encoder: it will first search the meta-path based neighbors $\mathcal{N}(v)$ for each node $v$, and then each node will attentively aggregate information from its neighbors. To learn the importance of the information from neighborhood nodes, we first present each relation type $r\in \mathcal R$ in the constructed AHIN by $\mathbf{R}_r\in \mathbb{R}^{d_{\mathbf{a}}\times d_{\mathbf{a}}}$, where $d_{\mathbf{a}}$ denotes the dimension of the attributed feature vector; and then the attentive weight $\beta$ of node $u$ (the neighbor of $v$) indicate the relevance of these two nodes measured in terms of the space $\mathbf R_r$, that is,
\begin{equation}\label{key0}
\beta_r(v,u)=\mathbf{a}_v^T\mathbf{R}_r\mathbf{a}_u,
\end{equation}
where $\mathbf{a}_v$ and $\mathbf{a}_u$ are the attributed feature vectors attached to node $v$ and $u$. We further normalize the weights across all the neighbors of $v$ by applying softmax function:
\begin{equation}
\widetilde{\beta}_r(v,u)=\dfrac{\exp(\beta_r(v,u))}{\sum_{u'\in \mathcal{N}(v)}\exp(\beta_r(v,u'))}.
\end{equation}
Then, the neighbors' representations can be formulated as the linear combination: 
\begin{equation}
\label{key1}\mathbf{a}_{\mathcal{N}(v)}=\widetilde{\beta}_r(v,u)\mathbf{a}_u,
\end{equation}
where the weight $\widetilde{\beta}_r(v,u)$ indicates the information propagated from $u$ to $v$ in terms of relation $r$. Finally, we aggregate $v$'s representation $\mathbf{a}_v$ and its neighbors' representations $\mathbf{a}_{\mathcal{N}(v)}$ by: 
\begin{equation}\label{eq:final}
\mathbf{a}_v=avg(\mathbf{a}_v+\mathbf{a}_{\mathcal{N}(v)}).
\end{equation}

\noindent \textit{Decoder.} The decoder is used to reconstruct the network topological structure. More specifically, based on the latent representations generated from the encoder, the decoder is trained to predict whether there is a link between two nodes in the constructed AHIN.

To this end, leveraging latent representations learned from the heterogeneous GAE, the risk index of a given area is calculated as:
\begin{equation}\label{key2}
Idx(v)=\sum_{i=1}^{d_{\mathbf{a}}}\gamma_i \mathbf{a}_v(i),
\end{equation}
where $\gamma_i$ is the adjustable parameter that can be specified by human experts, indicating the importance of $i$-th element in $\mathbf{a}_v$ (e.g., the number of confirmed cases, population density, age distribution, mobility measure, etc.) in the rapidly changing situation.

\section{System Development, Benchmark Datasets and Case Studies} \label{experiment}

Because of the critical need to act promptly and deliberately in this rapidly changing situation, we have deployed our developed system $\alpha$\textit{-Satellite} (i.e., an AI-driven system to automatically provide hierarchical community-level risk assessment related to COVID-19) for public test. Given a specific location (either user input or automatic positioning), the developed system will automatically provide risk indexes associated with it in a hierarchical manner (e.g., state, county, city, specific location) to enable people to select appropriate actions for protection while minimizing disruptions to daily life. The link of the system is: \textbf{https://COVID-19.yes-lab.org}, which also include the brief description and disclaimer of the system as well as the following benchmark datasets. 

\vspace{-0.2cm}
\subsection{Benchmark Datasets for Public Use}\label{subsec:experimentsetup}

\textbf{Data Collection and Preprocessing.} We have developed a set of crawling and preprocessing tools to collect and parse the large-scale and real-time pandemic related data from multiple sources, including disease related data from official public health organizations and digital media, demographic data, mobility data, and user generated data from social media (i.e., Reddit). We have made our collected and proprocessed data available for public use through the above link. We describe each publicly accessible benchmark dataset (i.e., $\mathbf{DB}_1$ - $\mathbf{DB}_4$) in detail below.  

\noindent \textbf{$\mathbf{DB}_1$: disease related dataset.} According to simplemaps\footnote{https://simplemaps.com/}, the U.S. includes 50 states, Washington, D.C. and Puerto Rico as well as 3,203 counties and 28,889 cities. We have collected the up-to-date county-based coronavirus related data including the numbers of confirmed cases, new cases, deaths and the fatality rate, from  official public health organizations (e.g., WHO, CDC, and county government websites) and digital media with real-time updates of COVID-19 (e.g., 1point3acres). By the date, we have collected these data from 1,531 counties and 52 states (including Washington, D.C. and Puerto Rico) on a daily basis from Feb. 28, 2020 to date (i.e., March 25, 2020). 

\noindent \textbf{$\mathbf{DB}_2$: demographic and mobility dataset.} We parse the demographic data collected from the the United States Census Bureau (data updated on July 1, 2019) in a hierarchical manner: for each city, county or state in the U.S., the dataset includes its estimated population, population density (e.g., number of people per square mile), age and gender distributions. By the date, we make the demographic and mobility dataset available for public use including the information of estimated population, population density, and GPS coordinates for 28,889 cities, 3,203 counties and 52 states (including Washington, D.C. and Puerto Rico). 

\noindent \textbf{$\mathbf{DB}_3$: social media data from Reddit.} In this work, we initialize our efforts on social media data with the focus of public perception analysis on Reddit, as it provides the platform for scientific discussion of dynamic policies, announcements, symptoms and events of COVID-19. In particular, we have collected and analyzed 48 state-based subreddits (i.e., Washington, D.C. and 47 states). By the date, we have crawled and automatically analyze 22,992 posts by 8,948 users on Reddit associated with 182,554 comments by 30,147 users on the discussion of COVID-19 from February 17, 2020 to date (i.e., March 25, 2020). Along with these data, this publicized dataset also includes the extracted locations of the posts using Stanford Named Entity Recognizer.

\noindent \textbf{$\mathbf{DB}_4$: constructed AHIN.} Based on our designed AHIN network schema in this work (shown in Figure \ref{fig:schema}), the constructed AHIN has 32,145 nodes (i.e., 1 node with type of nation, 52 nodes with type of state, 3,203 nodes with type of county, 28,889 nodes with type of city) and 96,459 edges (including 32,144 edges with relation type of \textit{R1} and 64,315 edges with relation type of \textit{R2}).

\vspace{-0.2cm}
\subsection{Case Studies}\label{subsec:exp1}

In this section, we evaluate the practical utility of the developed system $\alpha$\textit{-Satellite} for hierarchical community-level risk assessment related to COVID-19 through a set of case studies.

\begin{figure*}[ht]
	\centering
	\includegraphics[width=1\linewidth]{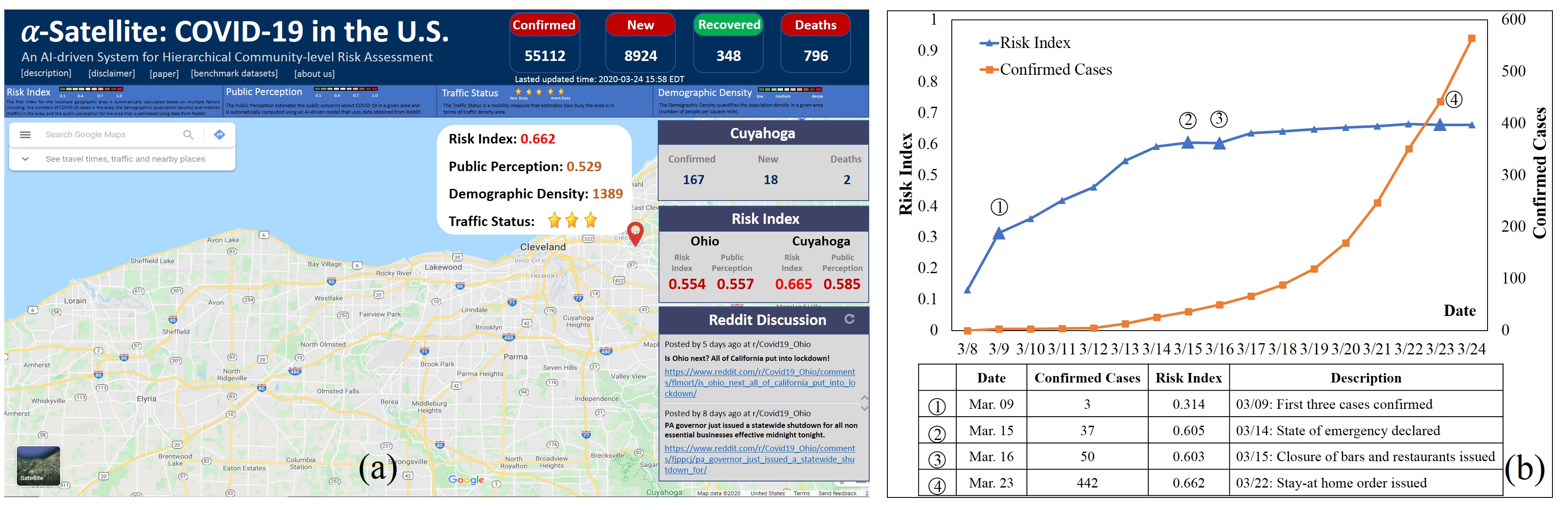}
	\caption{Risk index of a given area and comparisons of the indexes on different dates.}
	\label{fig:case1}
\end{figure*}

\noindent \textbf{Case study 1: real-time risk index of a given area.} Given a specific location (either user input or automatic positioning by Google map), the developed system will automatically provide its related risk index (i.e., ranging from [0,1], the larger number indicates higher risk and vice versa) associated with the public perceptions (i.e., awareness) towards COVID-19 in this area (i.e., ranging from [0,1], the larger number denotes more aware and vice versa), demographic density (i.e., the number of people per square mile in its related county), and traffic status (i.e., ranging from [1,5], the larger number means more traffic and vice versa). Figure \ref{fig:case1}.(a) shows an example: given the location of Euclid Ave, Cleveland, OH 44106, the risk index provided by the system was 0.662 (with public perception of 0.529, demographic density of 1,389, and traffic status of 3) at 3:58pm EDT on March 24, 2020.  At the same time, the risk indexes and public perceptions of corresponding county (i.e., Cuyahoga county with risk index of 0.665 and public perception of 0.585) and state (i.e., OH state with risk index of 0.554 and public perception of 0.557) will also be shown in a hierarchical manner to enable people to select appropriate actions for protection while minimizing disruptions to daily life.

\noindent \textbf{Case study 2: comparisons of risk indexes on different dates.} In this study, given the same area, we examine how the generated risk indexes change over time. Using the same location above, Figure \ref{fig:case1}.(b) shows the comparison results on different dates at the time of 3:58pm EDT, from which we have the following observations: 1) in general, its risk indexes increased over days from March 8, 2020 (i.e., 0.131) to March 24, 2020 (i.e., 0.662), as the confirmed cases in its related county (i.e., Cuyahoga county) and its related state (i.e., OH) continued to grow (i.e., from 0 case in Cuyahoga county on March 8 to 167 cases and 2 deaths on March 24, and from 0 case in OH on March 8 to 564 cases and 8 deaths on March 24); 2) after the first three case were confirmed in Cuyahoga county at OH on March 9, there was a sharp rise of risk index compared with March 8 (from 0.131 to 0.314); 3) the increases of risk indexes relatively slowed down after the public health and executive orders were issued in responses to COVID-19. For example, the risk indexes dropped to 0.605, 0.603 and 0.662 on March 15, 16 and 23 respectively, which might be because the government declared a state of emergency on March 14, ordered Ohio bars and restaurants to close on March 15 and issued a stay-at-home order on March 22\footnote{https://coronavirus.ohio.gov/wps/portal/gov/covid-19/home/public-health-orders/public-health-orders}.

\begin{figure*}[ht]
	\centering
	\includegraphics[width=1\linewidth]{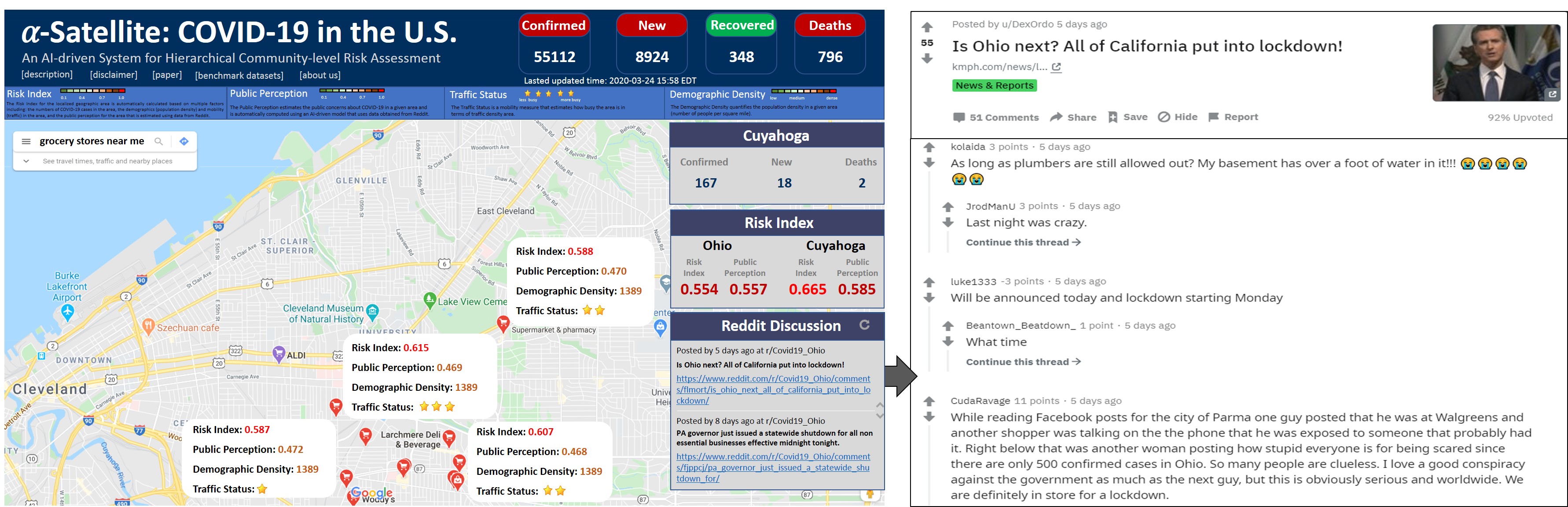}
	\caption{Comparisons of the indexes at different areas given the same time.}
	\label{fig:case2}
\end{figure*}

\noindent \textbf{Case study 3: comparisons of risk indexes at different areas.} In this study, given the same time, we examine how the generated risk indexes change over areas.  When a user inputs the areas he/she are interested in (e.g., \textit{grocery stores near me}) in the search bar, the system will display the nearby grocery stores using Google maps application programming interface (API) and automatically provide the associated indexes. For example, using the same time in the first study (i.e., 3:58pm EDT on March 24, 2020), Figure \ref{fig:case2} shows the ``grocery stores near me'' (i.e., near the location of Euclid Ave, Cleveland, OH 44106) and their related indexes. From Figure \ref{fig:case2}, we can observe that the indexes of nearby areas might vary due to the factors of different public perceptions towards COVID-19 and different traffic statuses in specific areas. As shown in the right part of Figure \ref{fig:case2}, the system also provides related Reddit posts to users.

\begin{figure*}[t]
	\centering
	\includegraphics[width=1\linewidth]{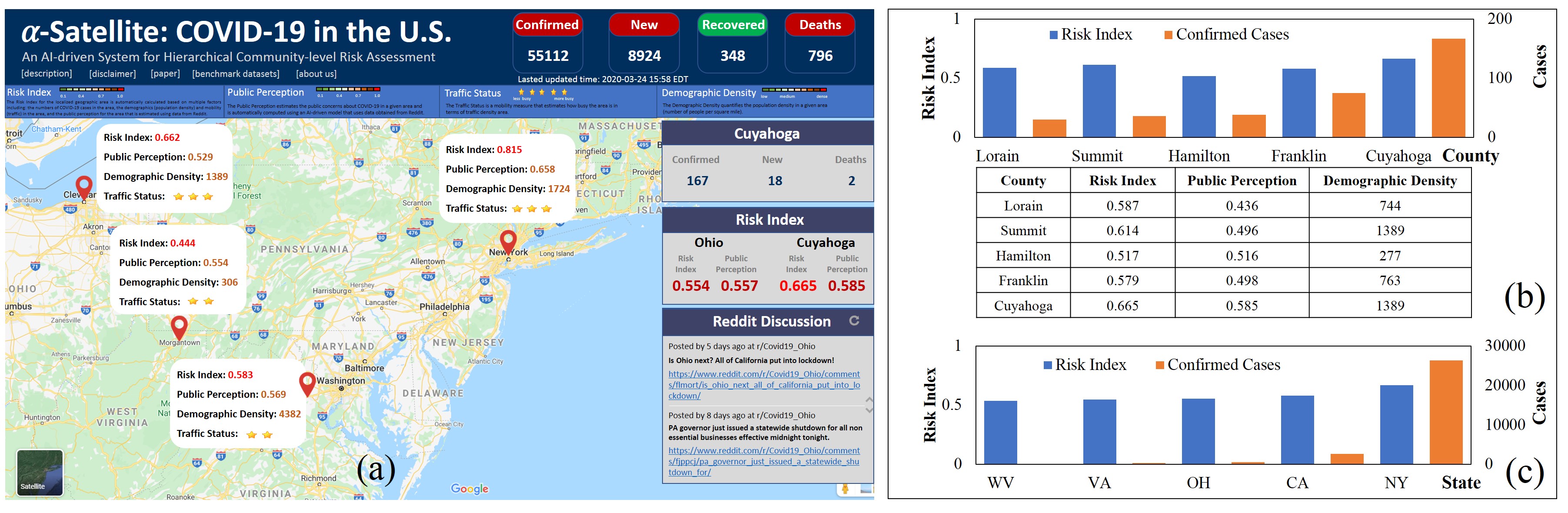}
	\caption{Comparisons of different counties and different states given the same time.}
	\label{fig:case3}
\end{figure*}

\noindent \textbf{Case study 4: comparisons of different counties and states.} In this study, we compare the indexes of different counties and different states given the same time. Using the time in the first study (i.e., 3:58pm EDT on March 24, 2020), Figure \ref{fig:case3}.(a) shows an example of comparisons. More specifically, at county-level, using OH state as an example, we choose the counties with top five largest numbers of confirmed cases on March 24 for comparisons: Cuyahoga (167), Franklin (75), Hamilton (38), Summit (36) and Lorain (30). Figure. \ref{fig:case3}.(b) illustrates the risk indexes associated with multiple factors versus the numbers of confirmed cases in these counties. For the comparisons of different states, we also choose five states: two most severe states (New York (NY) with 26,376 confirmed cases and 271 deaths, California (CA) with 2,628 confirmed cases and 54 deaths), two medium severe states (OH with 564 confirmed cases and 8 deaths, Virginia (VA) with 304 confirmed cases and 9 deaths) and one least severe state (West Virginia (WV) with 39 confirmed cases and 0 deaths). Figure. \ref{fig:case3}.(c) shows the risk indexes versus the numbers of confirmed cases in these states, from which we can see that there is a positive correlation between the numbers of confirmed cases and the risk indexes.

\section{Conclusion and Future Work} \label{conclusion}

To track the emerging dynamics of COVID-19 pandemic in the U.S., in this work, we propose to collect and model heterogeneous data from a variety of different sources, devise algorithms to use these data to train and update the models to estimate the spread of COVID-19 and predict the risks at community levels, and thus help provide actionable information to users for community mitigation. In sum, leveraging the large-scale and real-time data generated from heterogeneous sources, we have developed the prototype of an AI-driven system (named $\alpha$\textit{-Satellite})  to help combat the deadly COVID-19 pandemic. The developed system and generated benchmark datasets have made publicly accessible through our website. 

In the future work, we plan to continue our efforts to expand the data collection and enhance the system to help combat the fast evolving COVID-19 pandemic. We will continue to release our generated data and updates of the system to facilitate researchers and practitioners on the research to help combat COVID-19 pandemic, while assisting people to select appropriate actions to protect themselves at increased risk of COVID-19 while minimize disruptions to daily life to the extent possible.

\section{Acknowledgments} \label{ack}

Y. Ye, S. Hou, Y. Fan, Y. Qian, Y. Zhang, S. Sun and Q. Peng's work is partially supported by the NSF under grants IIS-1951504, CNS-1940859, CNS-1946327, CNS-1814825 and OAC-1940855, and by DoJ/NIJ under grant NIJ 2018-75-CX-0032. This work is also partially supported by the Institute for Smart, Secure and Connected Systems (ISSACS) at Case Western Reserve University.

\bibliographystyle{ACM-Reference-Format}
\bibliography{reference_SaTC}

\end{document}